\documentclass[aps,twocolumn,prl,preprintnumbers,amsmath,amssymb,superscriptaddress]{revtex4}
\usepackage{graphicx,epsfig}
\usepackage{dcolumn}
\usepackage{bm}
\usepackage{color}

\begin{document}

\title{Robust dynamic classes revealed by measuring the response
function of a social system}

\author{R. Crane}
\affiliation{Chair of Entrepreneurial Risks, Department of
Management, Technology and Economics, ETH-Zurich, CH-8001 Zurich,
Switzerland}

\author{D. Sornette}

\affiliation{Chair of Entrepreneurial Risks, Department of
Management, Technology and Economics, ETH-Zurich, CH-8001 Zurich,
Switzerland}

\date{\today}




\begin{abstract}
We study the relaxation response of a social system after endogenous
and exogenous bursts of activity using the time-series of daily
views for nearly 5 million videos on YouTube. We find that most
activity can be described accurately as a Poisson process. However,
we also find hundreds of thousands of examples in which a burst of
activity is followed by an ubiquitous power-law relaxation governing
the timing of views. We find that these relaxation exponents cluster
into three distinct classes, and allow for the classification of
collective human dynamics. This is consistent with an epidemic model
on a social network containing two ingredients:  A power law
distribution of waiting times between cause and action and an
epidemic cascade of actions becoming the cause of future actions.
This model is a conceptual extension of the fluctuation-dissipation
theorem \cite{ruelle2004fluctuation,roehner2004news} to social
systems, and provides a unique framework for the investigation of
timing in complex systems.

\end{abstract}

\maketitle




\section{Introduction}


Uncovering rules governing collective human behavior is a difficult
task because of the myriad of factors that influence an individual's
decision to take action.
Investigations into the timing of individual activity, as a basis
for understanding more complex collective behavior, have reported
statistical evidence that human actions range from random
\cite{haight1967poisson} to highly correlated
\cite{barabasi2005bursts}.
While most of the time the aggregated dynamics of our individual
activities create seasonal trends or simple patterns, sometimes our
collective action results in blockbusters, best-sellers, and other
large-scale trends in financial and cultural markets.

Here, we attempt to understand this nontrivial herding by
investigating how the distribution of waiting times describing
individuals' activity \cite{vazquez2006burstmodeling} is modified by
the combination of interactions \cite{vazquez2007interactions} and
external influences in a social network.
This is achieved by measuring the response function of a social
system \cite{roehner2004news} and distinguishing whether a burst of
activity was the result of a cumulative effect of small endogenous
factors or instead the response to a large exogenous perturbation.
%
%
Looking for endogenous and exogenous signatures in complex systems
provides a useful framework for understanding many complex systems
and has been successfully applied in several other contexts
\cite{sornette2005origins}.

\begin{figure}[h]
\centerline{\epsfig{figure=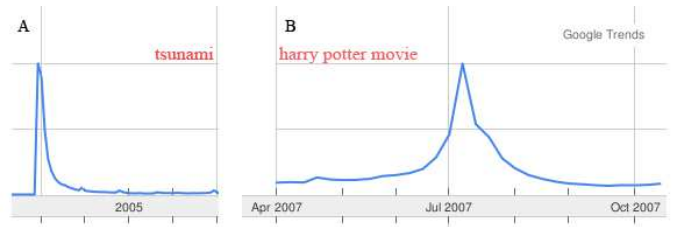,angle=0,width=9.0cm}}
\caption{ (color online) Search queries as a proxy for collective
human attention. (A) The volume of searches for the word ``tsunami''
in the aftermath of the catastrophic asian tsunami.  The sudden peak
and slow relaxation illustrate the typical signature of an
``exogenous'' burst of activity. (B) The volume of search queries
for ``Harry Potter movie''.  The significant growth preceding the
release of the film and symmetric relaxation is characteristic of an
``endogenous'' burst of activity.\label{fig:Google}}
\end{figure}

As an illustration of this distinction in a social system, consider
the example of trends in queries on internet search engines
\cite{google2008trends} in figure \ref{fig:Google}, which shows the
remarkable differences in the dynamic response of a social network
to major social events. For the ``exogenous'' catastrophic Asian
tsunami of December 26th, 2004, we see the social network responded
suddenly. In contrast, the search activity surrounding the release
of a Harry Potter movie has the more ``endogenous'' signature
generated by word-of-mouth, with significant precursory growth and
an almost symmetric decay of interest after the release.
In both ``endo'' and ``exo'' cases there is a significant burst of
activity. However, we expect to be able to distinguish the post-peak
relaxation dynamics on account of the very different processes that
resulted in the bursts.  Furthermore, we expect the relaxation
process to depend on the interest of the population since this will
influence the ease with which the activity can be spread from
generation to generation.

To translate this qualitative distinction into quantitative results,
we describe a model of epidemic spreading on a social network
\cite{sornette2003memory} and validate it with a data set that is
naturally structured to facilitate the separation of this endo/exo
dichotomy. Our data consists of nearly 5 million time-series of
human activity collected sub-daily over 8 months from the 4th most
visited web site (YouTube). At the simplest level, viewing activity
can occur one of three ways: randomly, exogenously (when a video is
featured), or endogenously (when a video is shared). This provides
us with a natural laboratory for distinguishing the effects that
various impacts have and allows us to measure the social ``response
function''.


\section{The model}

Various factors may lead to viewing a video, which include chance,
triggering from email, linking from external websites, discussion on
blogs, newspapers, and television, and from social influences. The
epidemic model we apply to the dynamics of viewing behavior on
YouTube uses two ingredients whose interplay capture these effects.

The first ingredient is a power law distribution of waiting times
describing human activity
\cite{barabasi2005bursts,vazquez2006burstmodeling,oliveira2005correspondence}
that expresses the latent impact of these various factors using a
response function which, on the basis of previous work
\cite{deschatres2005amazon,johansen2000internaut,johansen2001internaut},
we take to be a long-memory process of the form
\begin{equation}
\phi(t) \sim 1/t^{1+\theta}~,\ \ \ {\rm with}\ \ \  0 < \theta < 1~.
\label{eq:memorybare}
\end{equation}
By definition, the memory kernel $\phi(t)$ describes the
distribution of waiting times between ``cause'' and ``action'' for
an individual.  The ``cause'' can be any of the above mentioned
factors. The action is for the individual to view the video in
question after a time $t$ since she was first subjected to the
``cause'' {\it without} any other influences between $0$ and $t$,
corresponding to a direct (or first-generation) effect. In other
words, $\phi(t)$ is the ``bare'' memory kernel or propagator,
describing the direct influence of a factor that triggers the
individual to view the video in question. Here, the exponent
$\theta$ is the key parameter of the theory which will be determined
empirically from the data.

The second ingredient is an epidemic branching process that
describes the cascade of influences on the social network. This
process captures how previous attention from one individual can
spread to others and become the cause that triggers their future
attention \cite{sornette2004amazon}. In a highly connected network
of individuals whose interests make them susceptible to the given
video content, a given factor may trigger action through a cascade
of intermediate steps. Such an epidemic process can be conveniently
modeled by the so-called self-excited Hawkes conditional Poisson
process \cite{hawkes1974cluster}.  This gives the instantaneous rate
of views $\lambda(t)$ as
\begin{equation}
\lambda(t) = V(t) + \sum_{i, t_{i} \le t} \mu_{i}\phi{(t-t_{i})}
\label{eq:intensity}
\end{equation}
where $\mu_{i}$ is the number of potential viewers who will be
influenced directly over all future times after $t_{i}$ by person
$i$ who viewed a video at time $t_{i}$.  Thus, the existence of
well-connected individuals can be accounted for with large values of
$\mu_{i}$.  Lastly, $V(t)$ is the exogenous source, which captures
all spontaneous views that are not triggered by epidemic effects on
the network.

\subsubsection{Predictions of the model}

According to our model, the aggregated dynamics can be classified by
a combination of the type of disturbance (endo/exo) and the ability
of individuals to influence others to action
(critical/sub-critical), all of which is linked by a common value of
$\theta$. The following classification of behaviors emerges from the
interplay of the bare long-memory kernel $\phi(t)$ given by
(\ref{eq:memorybare}) and the epidemic influences across the network
modeled by the Hawkes process (\ref{eq:intensity})

\begin{itemize}
\item {\bf Exogenous sub-critical}.
When the network is not ``ripe'' (that is, when connectivity and
spreading propensity are relatively small), corresponding to the
case when the mean value  $\langle\mu_{i}\rangle$ of $\mu_{i}$ is
less than $1$, then the activity generated by an exogenous event at
time $t_c$ does not cascade beyond the first few generations, and
the activity is proportional to the direct (or ``bare'') memory
function $\phi(t-t_{c})$:
\begin{equation}
A_{bare}(t) \sim \frac{1}{(t-t_{c})^{1+\theta}}~. \label{eq:bare}
\end{equation}

\item  {\bf Exogenous critical}.
If instead the network is ``ripe'' for a particular video, i.e.,
$\langle\mu_{i}\rangle$ is close to 1, then the bare response is
renormalized as the spreading is propagated through many generations
of viewers influencing viewers influencing viewers, and the theory
predicts the activity to be described by \cite{sornette2003memory}:
\begin{equation}
A_{ex-c}(t)  \sim \frac{1}{(t-t_{c})^{1-\theta}}~.
 \label{eq:EXC}
\end{equation}

\item {\bf Endogenous critical}.
If in addition to being ``ripe'', the burst of activity is not the
result of an exogenous event, but is instead fueled by endogenous
(word-of-mouth) growth, the bare response is renormalized giving the
following time-dependence for the view count before and after the
peak of activity \cite{sornette2003memory}:
\begin{equation}
A_{en-c}(t)  \sim \frac{1}{|t-t_{c}|^{1-2\theta}} ~. \label{eq:ENC}
\end{equation}

\item {\bf Endogenous sub-critical}.
Here the response is largely driven by fluctuations, and not bursts
of activity.  We expect that many time-series in this class will
obey a simple stochastic process.
\begin{equation}
A_{en-sc}(t)  \sim \eta(t) ~. \label{eq:ENSC}
\end{equation}

\end{itemize}

\begin{figure}[h]
\centerline{\epsfig{figure=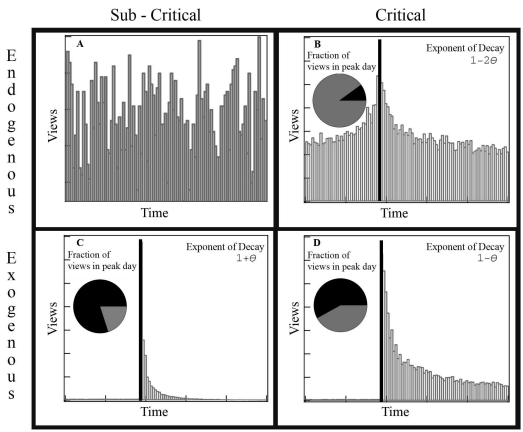,angle=0,width=7.0cm}}
\caption{A schematic view of the 4 categories described by our
models: (A) Endo-subcritical (B) Endo-critical (C) Exo-subcritical
(D) Exo-critical. The theory predicts the slope of the response
function conditioned on the class of the disturbance (endo/exo) and
the susceptibility of the network (critical/sub-critical).  Also
shown schematically in the pie chart is the fraction of views
contained in the main peak relative to the total number of views for
each category. This is used as a basis for sorting the time-series
into three distinct groups for further analysis of the exponents.
\label{fig:Predictions}}
\end{figure}

The dynamics described by the above classifications are illustrated
in figure \ref{fig:Predictions}. In addition to these classes, the
model predicts, by construction, a relationship between the fraction
of views obtained on the peak day compared to the total cumulative
views (figure \ref{fig:Predictions}: inset). For the
\textbf{exogenous sub-critical} class, the absence of precursory
growth and fast relaxation following a peak imply that close to
100\% of the views are contained in the peak. For the
\textbf{exogenous critical} class, the fractional views in the peak
should be smaller than the previous case on account of the content
penetrating the network resulting in a slower relaxation. Finally,
for the \textbf{endogenous critical} class, significant precursory
growth followed by a slow decay imply that the fractional weight of
this peak is very small compared to the total view count.  This
simple observation will be used as the basis for grouping exponents.

\section{Results}

\begin{figure}[h]
\centerline{\epsfig{figure=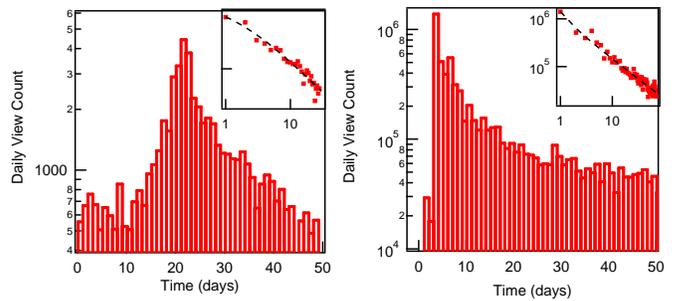,angle=0,width=9.0cm}}
\caption{(color online) An illustration of the typical response
found in hundreds of thousands of time-series on YouTube. (A) The
endogenous-critical class is marked by significant precursory growth
followed by an almost symmetric relaxation. (B) The
exogenous-critical class is marked by a sudden burst of activity
followed by a power-law relaxation. Inset shows the post-peak
relaxation on log-log scale, revealing the power-law behavior that
lasts over months for both classes.\label{fig:EndoExo}}
\end{figure}

We find that most videos' dynamics  ($\approx 90$\%) either do not
experience much activity or can be statistically described as a
random process (using a Poisson process and Chi-Squared test).  This
is not inconsistent with the endo-subcritical classification. For
the other 10\% ($\approx 500,000$ videos) we find nontrivial herding
behavior which accurately obeys the three power-law relations
described above. Characteristic examples of endogenous and exogenous
dynamics are shown in figure \ref{fig:EndoExo}.

For these videos that experience bursts, we calculate the fraction
$F$ of views on the most active day compared to the total count, and
define three classes:
\begin{enumerate}
\item Class 1 is defined by $80\% \leq F \leq 100 \%$.
\item Class 2 is defined by $20\% < F < 80\%$.
\item Class 3 is defined by $0\% \leq F \leq 20\%$.
\end{enumerate}

\begin{figure}[]
\centerline{\epsfig{figure=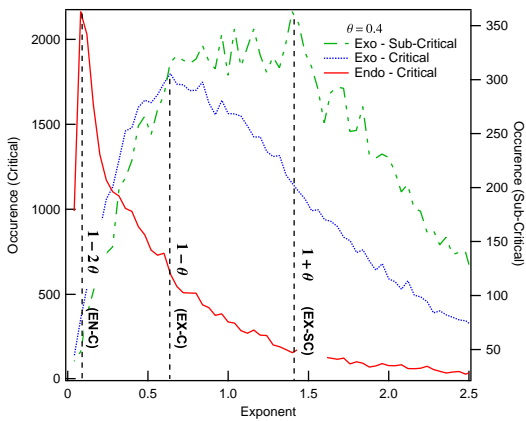,angle=0,width=7.0cm}}
\caption{(color online) Histogram of the exponents $p$  of the power
law relaxation $\sim 1/t^p$ of the view counts following a peak
belonging respectively to Class 1 (dashed green line), Class 2
(dotted blue line) and Class 3 (continuous red line). The predicted
values for the exogenous sub-critical class (equation
(\ref{eq:bare})), exogenous critical class (equation (\ref{eq:EXC}))
and for the endogenous critical class (equation (\ref{eq:ENC})) are
shown by the vertical dashed lines with their quantitative values
determined with the choice $\theta =0.4$. \label{fig:Histogram}}
\end{figure}

Should our model have any informative power, we should find a
correspondence
\begin{eqnarray}
{\rm Class\ 1}  & \longleftrightarrow &  {\rm Exogenous\ sub-critical}  \nonumber \\
{\rm Class\ 2}  & \longleftrightarrow &  {\rm Exogenous\ critical}  \nonumber \\
{\rm Class\ 3}  & \longleftrightarrow &  {\rm Endogenous\ critical}
\nonumber
\end{eqnarray}
Based on this simple classification, we show in figure
\ref{fig:Histogram} the histogram of exponents characterizing the
power law relaxation $\sim 1/t^{p}$. The exponents in the various
classes cluster into groups with the most probable exponent in each
class given respectively by $p \approx 1.4$, $0.6$, and $0.2$. These
results are robust with respect to changes of the threshold
percentages. These values are compatible with the predictions
(\ref{eq:bare}), (\ref{eq:EXC}), (\ref{eq:ENC}) of the epidemic
model with a unique value of $\theta = 0.4 \pm 0.1$.

\begin{figure}[]
\centerline{\epsfig{figure=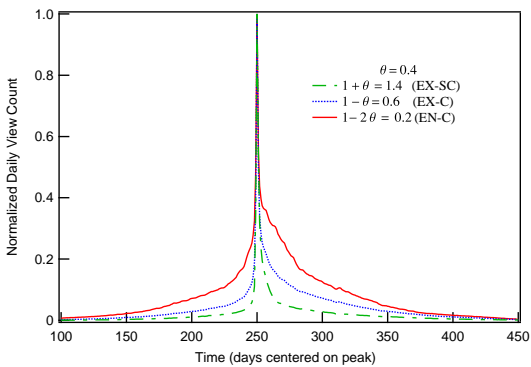,angle=0,width=7.0cm}}
\caption{(color online) Test of the precursory dynamics. The
(endo/exo) and (critical/subcritical) classification is based on
measuring the exponent governing the relaxation after a peak.
However, the epidemic branching model also predicts significant
precursory growth before a peak for the endogenous class with $p$
centered on $1-2 \theta \approx 0.2$ and no precursory growth for
the two exogenous classes.  The figure shows the peak-centered,
aggregate sum for all videos with exponents near either 1.4 (ex-sc),
0.6 (ex-c), or 0.2 (en-c).  We observe that videos classified as
endogenous-critical (continuous red line) on the basis of their
relaxation exponent, indeed have significantly more precursory
growth. One also sees very little precursory growth for the two
exogenous classes. \label{fig:Aggregate}}
\end{figure}

Having empirically extracted a value of $\theta$, a further test of
the model is provided by asking if the dynamics of videos with these
exponents are consistent with the description of the model. Here,
the test of the epidemic model lies in the precursory dynamics.  We
check this by performing a peak-centered, aggregate sum for all
videos with exponents near either 1.4, 0.6, or 0.2, with the intent
of visualizing the time evolution. Each time-series is first
normalized to 1 to avoid a single video from dominating the sum, and
the final result is divided by the number of videos in each set so
we can compare the three classes. The model predicts, and we indeed
observe in figure \ref{fig:Aggregate}, that videos whose post-peak
dynamics is governed by small exponents have significantly more
precursory growth. One also sees very little precursory growth for
the two exogenous classes.  Since by construction our grouping was
based on the exponent characterizing the relaxation after the peak,
one is not surprised to visualize faster decays for those videos
with exponents near 1.4 compared with those of 0.6 and 0.2.

\section{Discussion}

These results provide direct evidence that collective human dynamics
can be robustly classified by epidemic models, and understood as the
transformation of the distribution of individual waiting times by
exogenous and endogenous forces. One of the surprising results is
that the various classes are related by a common value of $\theta$.
While it is not expected that $\theta$ is universal in human
systems, it is similar to what has been found in other studies
\cite{sornette2004amazon}.  This provides a possible measure of the
strength of interactions in a social network, and will be the
subject of future work.

In addition to these results, understanding collective human
dynamics opens the possibility for a number of tantalizing
applications. It is natural to suggest a qualitative labeling that
is quantitatively consistent with the three classes derived from the
model: viral, quality, and junk videos. Viral videos are those with
precursory word-of-mouth growth resulting from epidemic-like
propagation through a social network and correspond to the
endogenous critical class with an exponent ($1-2\theta$). Quality
videos are similar to viral videos but experience a sudden burst of
activity rather than a bottom-up growth. Because of the ``quality''
of their content they subsequently trigger an epidemic cascade
through the social network. These correspond to the exogenous
critical class, relaxing with an exponent ($1-\theta$). Lastly, junk
videos are those that experience a burst of activity for some reason
(spam, chance, etc) but do not spread through the social network.
Therefore their activity is determined largely by the
first-generation of viewers, corresponding to the exogenous
sub-critical class, and they should relax as ($1+\theta$). While one
might argue that these labels are inherently subjective, they
reflect the objective measure contained in the collective response
to events and information.  This is further supported by the average
number of total views in each class, which is largest for ``viral''
(33,693 views) and smallest for ``junk'' (16,524 views) as one would
expect.

While the above description applies to videos, one could extend this
technique to the realm of books
\cite{deschatres2005amazon,sornette2004amazon}, movies, and other
commercial products, perhaps using sales as a proxy for measuring
the relaxation of individual activity. The proposed method for
classifying content has the important advantage of robustness as it
does not rely on qualitative judgment, using information revealed by
the dynamics of the human activity as the referee. More importantly,
the method does not rely on the magnitude of the response because of
the scale free nature of the relaxation dynamics. This implies that
identification of relevance---or lack of relevance---can be made for
content that has mass-appeal, along with that which appeals to more
specialized communities.  Furthermore, this framework could be used
to provide a quantitative measure of the effectiveness of marketing
campaigns by measuring the sales response to an ``exogenous''
marketing event.

A tenant of complex systems theory is that many seemingly disparate
and unrelated systems actually share an underlying universal
behavior.  In the digital age, we now have access to unprecedented
stores of data on human activity.  This data is usually almost
trivial to acquire---in both time and money---when compared with
'traditional'  measurements. If the complex behavior in social
systems is shared by other complex systems, then our approach, which
disentangles the individual response from the collective, may
provide a useful framework for the study of their dynamics.



\begin{acknowledgments}
We would like to thank Gilles Daniel, Ryan Woodard, Georges Harras,
N. Peter Armitage, and Yannick Malvergne for invaluable discussions
and comments on the manuscript.
\end{acknowledgments}




\end{document}